\documentclass[prd,twocolumn,amssymb,amsmath,nofootinbib]{revtex4-1}

\usepackage{color}
\definecolor{linkcolor}{RGB}{181,101,167} % color of the year!
\definecolor{urlcolor}{RGB}{97,66,151}
\usepackage[colorlinks=true
,urlcolor=urlcolor
,anchorcolor=linkcolor
,citecolor=linkcolor
,filecolor=linkcolor
,linkcolor=linkcolor
,menucolor=linkcolor
,linktocpage=true
,pdfproducer=medialab
]{hyperref}

\usepackage{graphicx}

\makeatletter
\newsavebox\myboxA
\newsavebox\myboxB
\newlength\mylenA

\newcommand*\xoverline[2][0.75]{%
    \sbox{\myboxA}{$\m@th#2$}%
    \setbox\myboxB\null% Phantom box
    \ht\myboxB=\ht\myboxA%
    \dp\myboxB=\dp\myboxA%
    \wd\myboxB=#1\wd\myboxA% Scale phantom
    \sbox\myboxB{$\m@th\overline{\copy\myboxB}$}%  Overlined phantom
    \setlength\mylenA{\the\wd\myboxA}%   calc width diff
    \addtolength\mylenA{-\the\wd\myboxB}%
    \ifdim\wd\myboxB<\wd\myboxA%
       \rlap{\hskip 0.5\mylenA\usebox\myboxB}{\usebox\myboxA}%
    \else
        \hskip -0.5\mylenA\rlap{\usebox\myboxA}{\hskip 0.5\mylenA\usebox\myboxB}%
    \fi}
\makeatother

\newcommand{\cD}{\mathcal{D}}

\newcommand{\cL}{\mathcal{L}}

\newcommand{\cN}{\mathcal{N}}
\newcommand{\cO}{\mathcal{O}}

\newcommand{\cV}{\mathcal{V}}

\newcommand{\uD}{\mathrm{D}}

\newcommand{\us}{\mathrm{s}}

\newcommand{\ga}{\gamma}
\newcommand{\ka}{\kappa}

\newcommand{\vt}{\vartheta}

\newcommand{\np}{\mathrm{np}}
\newcommand{\Pl}{\mathrm{pl}}

\newcommand{\Mpl}{M_{\mathrm{pl}}}

\newcommand{\SU}[1]{\mathrm{SU}\!\left(#1\right)}

\newcommand{\magf}{F}
\newcommand{\windnum}{\aleph}

\begin{document}

\title{Aligned Natural Inflation in String Theory}

\author{Cody Long}
\email{cel89@cornell.edu}
\author{Liam McAllister}
\email{mcallister@cornell.edu}
\author{Paul McGuirk}
\email{mcguirk@cornell.edu}

\affiliation{Department of Physics, Cornell University, Ithaca, New
  York, 14853, USA}

\begin{abstract}
We propose a scenario for realizing super-Planckian axion decay
constants in Calabi-Yau orientifolds of type IIB string theory,
leading to large-field inflation.  Our construction is a simple
embedding in string theory of the mechanism of Kim, Nilles, and
Peloso, in which a large effective decay constant arises from alignment
of two smaller decay constants.  The key ingredient is gaugino
condensation on magnetized or multiply-wound D7-branes.  We argue
that, under very mild assumptions about the topology of the
Calabi-Yau, there are controllable points in moduli space with large
effective decay constants.
\end{abstract}

\maketitle

%%%%
%%%%
\section{Introduction}
%%%%
%%%%

Observations of the cosmic microwave background
(CMB)~\cite{Hinshaw:2012aka,*Ade:2013uln} have provided extraordinary
support for the inflationary paradigm of an early epoch of accelerated
cosmic expansion~\cite{Guth:1980zm,*Linde:1981mu,*Albrecht:1982wi}.
Recent measurements of B-mode polarization by the BICEP2
collaboration~\cite{Ade:2014xna} suggest that, in addition to the
scalar fluctuations that had been previously observed and
well-studied, the inflationary period was marked by significant tensor
fluctuations. In particular, if the observed polarization is
cosmological in origin, it is compatible with a tensor-to-scalar ratio
of $r\approx 0.1-0.2$.
 
Detectable primordial B-mode polarization implies that the energy
scale of inflation is comparable to the unification scale,
\begin{equation}
  V_{\mathrm{inf}}^{1/4}\approx 2.2\times 10^{16}\ \mathrm{GeV}\left(\frac{r}{0.2}\right)^{1/4}.
\end{equation}
In the simplest scenarios, inflation is driven by a single scalar
field $\phi$ undergoing slow-roll evolution in a potential $V$.  In
such models, the tensor-to-scalar ratio respects the Lyth
bound~\cite{Lyth:1996im}
\begin{equation}
  \frac{\Delta\phi}{M_{\mathrm{pl}}}\gtrsim
  \left(\frac{r}{0.01}\right)^{1/2},
\end{equation}
in which $\Delta \phi$ is the distance in field space that the
inflaton moves during inflation and $M_{\mathrm{pl}}\approx 2.4\times
10^{18}\ \mathrm{GeV}$ is the reduced Planck scale.  The BICEP2
observation, taken at face value, implies a trans-Planckian displacement in field space for
such an inflaton. Analogous bounds exist for theories with multiple
fields or non-canonical kinetic terms, with the
general result that the observed $r$ requires $\Delta\phi$ to exceed
the ultraviolet cutoff of the theory~\cite{Baumann:2011ws}.

Inflation is famously sensitive to corrections from quantum gravity.
Even if the scale of inflation were much smaller than
$M_{\mathrm{pl}}$, Planck-suppressed operators can make
$\cO\left(1\right)$ corrections to the slow-roll parameter
$\eta=M_{\Pl}^{2}\frac{V''}{V}$, which must remain small during
inflation.  In addition to this universal $\eta$ problem, the large
inflationary scale and the trans-Planckian excursion of the inflaton
suggested by the BICEP2 results intensify the need for a consistent
embedding of inflation into a theory of quantum gravity such as string
theory.\footnote{See~\cite{Baumann:2014nda} for a recent comprehensive
  review of inflationary models in string theory.}  However, despite
much progress, complete and explicit models of large-field inflation
in string theory remain elusive.

In this note, we suggest a mechanism by which large-field inflation
may be realized in controllable string theory constructions.  We make
use of the scheme of natural inflation~\cite{Freese:1990rb} and the
{\it{decay constant alignment}} mechanism proposed by Kim, Nilles, and
Peloso (KNP)~\cite{Kim:2004rp}.\footnote{See~\cite{Kappl:2014lra, Tye:2014tja,
    Czerny:2014xja, *Czerny:2014qqa,*Higaki:2014pja} for recent
  field-theoretic treatments of related scenarios.} We
demonstrate that compactifications of type IIB string theory allow for
the large axion decay constants required to realize large-field
natural inflation in this framework.  However, we do not stabilize
moduli in this note, and so fall short of a complete model of
inflation.

While this work was in the last stages of preparation, we learned
of the independent concurrent development of closely related ideas by
other authors~\cite{Ben-Dayan:2014lca}.

%%%
%%%
\section{\label{sec:KNP}Natural Inflation and Decay Constant Alignment}
%%%
%%%

A natural way to control the appearance of irrelevant operators in the
inflationary potential, and thus ensure that $\eta$ is small, is to
suppose that the inflaton enjoys a global symmetry. The natural
inflation scenario~\cite{Freese:1990rb} achieves this by using an
axion to drive inflation.  At the classical level, an axion enjoys a
continuous shift symmetry $\phi\to\phi+c$ for arbitrary $c$. However,
when non-perturbative quantum corrections are taken into account, this
symmetry is broken to a discrete shift symmetry $\phi\to\phi+2\pi f$,
where $f$ is known as the axion decay constant.  An example of a 4d
Lagrangian that respects this symmetry is
\begin{equation}
  \cL=-\frac{1}{2}\bigl(\partial\phi\bigr)^{2}-\Lambda^{4}
  \biggl[1-\cos\frac{\phi}{f}\biggr].
\end{equation}
When $f\gtrsim 10\, M_{Pl}$ and $\Lambda\sim 10^{-3}\, \Mpl$, this
potential supports realistic inflation.  Indeed, for large $f$, the
potential is well-approximated by a quadratic potential,
$V\approx\frac{\Lambda^{4}}{2f^{2}}\phi^{2}$, while higher-order terms
in the potential are suppressed by $f$, rather than by $M_{\Pl}$, thus
ensuring the smallness of $\eta$.

It is of obvious importance to determine whether the super-Planckian
decay constants required to obtain realistic natural inflation can be
obtained in a theory of quantum gravity such as string theory.  String
theory compactifications can provide many axions descending from the
various $p$-form potentials of ten-dimensional supergravity
(see~\cite{Svrcek:2006yi} for a comprehensive discussion), but there
exist no complete examples in which the decay constants are suitably
large and the construction is parametrically controlled.\footnote{See,
  however, \cite{Blumenhagen:2014gta,*Grimm:2014vva} for recent
  proposals to obtain large decay constants in string theory.}
Indeed, there are generic arguments that suggest that it is quite
difficult to obtain large decay constants in which the string
construction is under good control~\cite{Banks:2003sx}. However, it
was pointed out by KNP~\cite{Kim:2004rp} that if there are multiple
axions and multiple non-perturbative effects that couple to different
linear combinations of the axions, then an effectively large axion
decay constant can be obtained even when the original decay constants
are modest.  Such an enhancement relies on the alignment of the decay
constants appearing in the non-perturbative corrections to the scalar
potential.

To illustrate this alignment, we follow~\cite{Kim:2004rp} and consider
a theory with two axions $\phi^{1}$ and $\phi^{2}$ with the
non-perturbatively generated potential
\begin{align}
  V\bigl(\phi\bigr)=\phantom{+}&\Lambda_{A}^{4}\biggl[1-\cos\biggl(
    \frac{\phi^{1}}{f_{A1}}+\frac{\phi^{2}}{f_{A2}}\biggr)\biggr]\notag\\
   +&\Lambda_{B}^{2}\biggl[1-\cos\biggl(
      \frac{\phi^{1}}{f_{B1}}+\frac{\phi^{2}}{f_{B2}}\biggr)\biggr] 
   \label{eq:KNP_pot}
\end{align} and the kinetic term \begin{equation}
  \cL_{\mathrm{kin}}=-k_{ij}\partial_{\mu}\phi^{i}\partial^{\mu}\phi^{j}.
\end{equation}
In general, the axions may have kinetic mixing as well, with
off-diagonal terms in $k_{ij}$.

Taking both the kinetic and potential mixings into account, the
determinant of the mass matrix of the canonically normalized fields is
\begin{equation}
  \label{eq:KNP_Hess_det}
  \left.\det \partial_{i}\partial_{j}V\right\vert_{\phi_{1}=\phi_{2}=0}
  =\frac{\left(f_{A1}f_{B2}-f_{A2}f_{B1}\right)^{2}}
  {f_{A1}^{2}f_{A2}^{2}f_{B1}^{2}f_{B2}^{2}}
  \frac{\Lambda_{A}^{4}\Lambda_{B}^{4}}
       {4\det\left(k\right)}
\end{equation}
and so a flat direction emerges when the decay constants are aligned
\begin{equation}
  \label{eq:alignment}
  \frac{f_{A1}}{f_{A2}}=\frac{f_{B1}}{f_{B2}}.
\end{equation}
When this alignment is achieved, both terms in~\eqref{eq:KNP_pot}
couple to the same linear combination of the axions, while the
orthogonal direction remains a flat direction.

By slightly misaligning the decay constants, a nearly flat direction
emerges and a large effective decay constant can be obtained for this
direction~\cite{Kim:2004rp}.  To illustrate this, we consider an
example of canonically normalized
axions\footnote{From~\eqref{eq:KNP_Hess_det}, it is apparent that the
  effect of introducing kinetic mixing (which generally decreases
  $\det\left(k\right)$) is to further lift the nearly flat direction
  resulting from near alignment.} for which all of the decay constants
are nearly equal while the dynamical scales are hierarchical
\begin{align}
  \label{eq:KNP_scaling_example}
  f_{A1}=f_{B1}=&f_{A2}=f,\quad
  f_{B2}=f\bigl(1+\delta\bigr),\\
  \Lambda_{A}^{4}=&\Lambda^{4},\quad
  \Lambda_{B}^{4}=\delta^{p}\Lambda^{4},
\end{align}
where $p>0$ and $\delta\ll 1$.  Then the determinant of the Hessian at the minimum is
\begin{equation}
  \left.\det \partial_{i}\partial_{j}V\right\rvert_{\phi=0}=
  \frac{\delta^{2+p}\Lambda^{8}}
  {f^{4}\left(1+\delta\right)^{2}},
\end{equation}
while the particular eigenvalues are
\begin{equation}
  m_{1}^{2}=\delta^{2+p}\frac{\Lambda^{4}}{2f^{2}}
  \left[1+\cO\left(\delta\right)\right]
  \qquad
  m_{2}^{2}=\frac{2\Lambda^{4}}{f^{2}}\left[1+\cO\left(\delta\right)\right].
\end{equation}
In terms of the mass eigenstates, the potential takes the form
\begin{multline}
  V=\Lambda^{4}\bigg[1-\cos\biggl(\frac{\sqrt{2}\psi^{2}}{f}
    +\frac{\delta^{p+1}\psi^{1}}{\sqrt{2}f}\biggr)\biggr]\\
  +\delta^{p}\,\Lambda^{4}
  \biggl[1-\cos\biggl(-\frac{\sqrt{2}\psi^{2}}{f}+
    \frac{\delta\psi^{1}}{\sqrt{2}f}\biggr)\biggr],
\end{multline}
where in each cosine we have kept only the first term in which either
field appears. The first condensate serves to stabilize
$\psi^{2}\approx 0$, while for sufficiently small $\psi^{1}$, the
potential is dominated by the second term, which exhibits an effective
decay constant that is parametrically enhanced by the near alignment:
\begin{equation}
  f_{\mathrm{eff}}\sim\frac{f}{\delta}.
\end{equation}
Interestingly, the first term in the condensate naively exhibits an
even larger effective decay constant.  However, its contribution to
the potential of $\psi^{1}$ (for fixed $\psi^{2}$) is subdominant for
small $\psi^{1}$.

Although we illustrated the effectiveness of decay constants with a
particular scaling~\eqref{eq:KNP_scaling_example}, the scheme works
more generally.  Indeed, one of the examples of an inflationary
potential that we present in \S\ref{sec:examples} utilizes different
relationships between the decay constants and the dynamical scales.

%%%%%
%%%%%
\section{\label{sec:axions}Axions in Type IIB String Theory}
%%%%%
%%%%%

Our objective in this note is to propose a framework in which the
success of decay constant alignment may be embedded into a string
compactification.  In general, the construction of inflationary models
in string theory cannot be decoupled from the problem of moduli
stabilization. Therefore, we focus on inflation in O3/O7 Calabi-Yau
orientifold compactifications of type IIB string theory, where the
understanding of moduli stabilization is presently the most mature.
Although we will not stabilize moduli, the ingredients that we use to
construct the inflationary potential are the same as those used for
moduli stabilization in this corner of the landscape, and we have
found no reason why a completely stabilized compactification could not
in principle be constructed. In this section, we briefly review
aspects of the effective field theories of such orientifolds, focusing
on elements that are relevant for the construction of our inflationary
potentials.  A more detailed treatment can be found in, for
example,~\cite{Grimm:2004uq}.

The IIB supergravity multiplet in ten dimensions consists of the metric, the axiodilaton
$\tau=C+ie^{-\Phi}$, the $2$-form potentials $B_{2}$ and $C_{2}$, and
the $4$-form potential $C_{4}$.  In the absence of sources, the
low-energy description of a compactification on a Calabi-Yau threefold
with Hodge numbers $\left(h^{1,1},h^{2,1}\right)$ is a
4d $\cN=2$ supergravity theory with $h^{2,1}$ vector
multiplets (the scalar components of which are the complex structure
moduli), $h^{1,1}$ hypermultiplets (including K\"ahler moduli), and
the universal hypermultiplet built from $\tau$ and $B_{\mu\nu}-\tau
C_{\mu\nu}$.

Breaking to $\cN=1$ in four dimensions can be accomplished by
orientifolding, in which we identify states related by an orientation
reversal of the worldsheet and a holomorphic involution of the
Calabi-Yau geometry.  The cohomology groups split under the action of
the involution
\begin{equation}
  H^{\left(p,q\right)}=H_{+}^{\left(p,q\right)}\oplus H_{-}^{\left(p,q\right)}.
\end{equation}
Correspondingly,
\begin{equation}
  h^{p,q}=h_{+}^{p,q}+h_{-}^{p,q},\quad h^{p,q}_{\pm}=\dim\,H_{\pm}^{\left(p,q\right)}.
\end{equation}
The geometric involution will have fixed loci corresponding to the
presence of orientifold planes.  We will focus on O3/O7 orientifolds
in which the fixed loci are points (O3-planes) and divisors
(O7-planes).  After the orientifold action, the low-energy theory is a
4d $\cN=1$ supergravity theory with $h_{+}^{2,1}$ vector multiplets
coming from $C_{4}$, $h_{-}^{2,1}$ chiral multiplets describing
complex structure deformations, a chiral multiplet with scalar
component $\tau$, $h_{-}^{1,1}$ chiral multiplets from the $2$-form
potentials, and $h_{+}^{1,1}$ chiral multiplets corresponding to
complexified K\"ahler moduli.  Finally, the orientifold planes carry
D3- and D7-brane charge that must be canceled by the inclusion of
D-branes. These give rise to additional low-energy degrees of freedom
corresponding to the deformations of these branes.

The axiodilaton, complex structure moduli, and the deformation moduli
of $\uD 7$-branes can be stabilized at the perturbative level by
fluxes.  We will assume that such fluxes have been included and that
these fields are stabilized at a high scale and can be consistently
integrated out.\footnote{Integrating out fields stabilized by fluxes
  is consistent in compactifications allowing sufficiently large
  hierarchies of scales, but for constructions of high-scale
  inflation, which allow only modest hierarchies, it would be
  worthwhile to relax this assumption.}  The remaining closed-string
moduli are those coming from the $2$-form potentials and the
complexified K\"ahler moduli.  Their scalar degrees of freedom are
encoded in the K\"ahler coordinates~\cite{Grimm:2004uq}
\begin{align}
  G^{a}=&c^{a}-\tau\,b^{a},\\
  T^{\alpha}=&\tau^{\alpha}+i\vt^{\alpha}
  +\frac{1}{4}e^{\phi}\ka^{\alpha}_{\phantom{\alpha}bc}
  G^{b}\left(G^{c}-\bar{G}^{\bar{c}}\right),
  \label{eq:T_def}
\end{align}
in which $a=1,\ldots,h_{-}^{1,1}$, $\alpha=1,\ldots,h_{+}^{1,1}$, and
\begin{align}
  \tau^{\alpha}+i\vt^{\alpha}&=\frac{1}{2}\int_{\cD^{\alpha}}J\wedge
  J
  +i\int_{\cD^{\alpha}}C_{4},\\
  C_{2}&=c^{a}\omega_{a},\quad
  B_{2}=b^{a}\omega_{a}.
\end{align}
The scalar field $\tau^{\alpha}$ is the volume of the $4$-cycle
$\cD^{\alpha}$, where $\cD^{\alpha}$ ($\cD_{a}$) is a
divisor that is even (odd) under the geometric involution and
$\omega^{\alpha}$ ($\omega_{a}$) is the Poincar\'e dual of an even
(odd) divisor. Here $\ka^{ijk}$ are triple intersection numbers in the
`upstairs' Calabi-Yau, and indices on the intersection numbers are
raised and lowered with the identity matrix.   Note that the
definition of the good K\"ahler coordinates is modified by the
presence of open
strings~\cite{Jockers:2004yj,*Jockers:2005pn,Jockers:2005zy} or strong
warping~\cite{Frey:2008xw,*Marchesano:2008rg,*Martucci:2009sf}.

At the classical level and in the absence of localized brane sources,
each of the fields $\vt^{\alpha}$, $c^{a}$, and $b^{a}$ enjoys a
continuous shift symmetry.  This is reflected by the absence of a
superpotential for $T^{\alpha}$, and a K\"ahler potential for the
K\"ahler moduli that depends only on 2-cycle volumes
\begin{equation}
  K=-2\log\,\cV,\quad
  \cV=\frac{1}{3!}\ka^{\alpha\beta\ga}t_{\alpha}t_{\beta}t_{\ga}.
\end{equation}
The volumes of 4-cycle volumes $\tau^{\alpha}$ are related to 2-cycle
volumes $t_{\alpha}$ via
$\tau^{\alpha}=\frac{1}{2}\ka^{\alpha\beta\ga}t_{\beta}t_{\ga}$ and
hence the K\"ahler potential is an implicit function of the K\"ahler
coordinates $T^{\alpha}$, $G^{a}$, and their conjugates.

Non-perturbatively, the continuous shift symmetries of $\vt^{\alpha}$,
$c^{a}$, and $b^{a}$ are broken to discrete shift symmetries.  For
example, a stack of $\uD 7$-branes realizing an $\SU{N}$ gauge theory
and wrapping a divisor $\cD$ contributes a non-perturbative correction
to the superpotential via gaugino condensation
\begin{equation}
  \label{eq:W_np}
   W_{\np}=Ae^{-aT},
\end{equation}
in which $T=\tau+\cdots$ corresponds to the volume of $\cD$, $A$ is a
function of the stabilized complex structure and brane moduli, and
$a=\frac{2\pi}{N}$.  In order for these superpotential terms to
appear, the deformation moduli of the $\uD 7$-branes must be lifted
either by flux or by taking the stack to wrap a rigid divisor.  The
coordinate $T$ appears because the tree-level gauge kinetic function
for the gauge theory realized by the D7-branes is simply
\begin{equation}
  \mathfrak{f}_{\uD 7}=T.
\end{equation}
Although the odd moduli $G^{a}$ do not appear in $W_{\np}$, the appearance of
$b^{a}$ in the real part of the K\"ahler coordinate $T^{\alpha}$ means
that the $b^{a}$ can be stabilized when the 4-cycle
volumes are stabilized
non-perturbatively~\cite{McAllister:2008hb,Flauger:2009ab}.  At this
level, the $c^{a}$ axions remain unstabilized, though non-perturbative
corrections to the K\"ahler potential from Euclidean D1-branes will generically
induce a mass for these fields.

The situation changes when the stack of $\uD 7$-branes is magnetized
or is allowed to have multiple windings.  We consider a restricted class of
magnetizations that can be expanded in terms of pullbacks of
$2$-forms $\omega_{a}$ on the Calabi-Yau
\begin{equation}
  \label{eq:wv_flux}
  F^{\mathrm{int}}_{2}=
  \frac{1}{2\pi\alpha'}\,\magf^{a}\mathrm{P}\bigl[\omega_{a}\bigr].
\end{equation}
When the stack is magnetized, the gauge kinetic function is modified
and depends holomorphically on the odd moduli $G^{a}$.  To be more
precise, let $\cD$ be the divisor wrapped by the D7-brane stack, and
let $\cD'$ be the orientifold image.  Then define the even and odd
cycles
\begin{equation}
  \cD^{\pm}=\cD\cup \bigl(\pm \cD'\bigr)
\end{equation}
and the wrapping numbers  
\begin{equation}
  \label{eq:wrapping_numbers}
  \windnum_{\alpha}=\int_{\cD^{+}}\tilde{\omega}_{\alpha},\qquad
  \windnum^{a}=\int_{\cD^{-}}\tilde{\omega}^{a},
\end{equation}
where $\tilde{\omega}$ are 4-forms satisfying
\begin{equation}
  \int\omega^{\alpha}\wedge\tilde{\omega}_{\beta}=\delta_{\beta}^{~\alpha},
  \qquad
  \int\omega_{a}\wedge\tilde{\omega}^{b}=\delta^{~b}_{a},
\end{equation}
in which the integral is taken over the Calabi-Yau.  The gauge kinetic
function for the D7-brane gauge theory is 
then~\cite{Jockers:2004yj,*Jockers:2005pn,Jockers:2005zy,Grimm:2011dj}
\begin{equation}
  \label{eq:gkf}
  \mathfrak{f}_{\uD 7}=\windnum_{\alpha}\biggl[T^{\alpha}+i\,\ka^{\alpha}_{\phantom{\alpha}bc}
  \biggl(G^{b}\magf^{c}+\frac{\tau}{2}\magf^{b}\magf^{c}\biggr)\biggr].
\end{equation}
The contribution to the superpotential for such a stack of magnetized
$\uD 7$-branes is $Ae^{-2\pi \mathfrak{f}_{\uD 7}/N}$ for an $\SU{N}$ gauge
theory.  Including such magnetization thus breaks the continuous shift
symmetry of $c^{a}$ to a discrete shift symmetry at the level of the
superpotential.  The shift symmetry for $b^{a}$ is badly broken by the
appearance of $b^{a}b^{a}$ in the real part of the K\"ahler 
moduli~\eqref{eq:T_def}.

The magnetization also contributes to the D-term for the $\uD 7$-brane gauge
theory~\cite{Jockers:2005zy,Grimm:2011dj}
\begin{equation}
  \label{eq:D-term}
  D_{\uD 7}=\frac{\alpha' t_{\alpha}}{2\cV}
  \ka^{\alpha}_{\phantom{\alpha}bc}\bigl(b^{b}-\magf^{b}\bigr)\windnum^{c}.
\end{equation}
In general, the D-term receives contributions from the matter fields
living on the D7-branes, but, as with the complex structure moduli and
the axiodilaton, we will assume that they have been stabilized at a
very high scale by closed-string flux.

%%%%%
%%%%%
\section{\label{sec:mag}Large Decay Constants from Magnetized Branes}
%%%%%
%%%%%

We will consider two different, though closely related, mechanisms by
which large axion decay constants can be realized via an
implementation of the alignment scenario reviewed in \S\ref{sec:KNP}.
In this section, we will show how such an alignment can be arranged by
the magnetization of homologous $\uD 7$-branes.\footnote{The wrapping
  of homologous but distinct cycles was utilized in related
  constructions of axion monodromy
  inflation~\cite{McAllister:2008hb}.}

As discussed in the previous section, in the presence of condensing
magnetized branes, $\vt^{\alpha}$ and $c^{a}$ enjoy discrete shift
symmetries, while $\tau^{\alpha}$ and $b^{a}$ do not.  Therefore,
$\vt^{\alpha}$ and $c^{a}$ are candidates for natural inflation.
Unless we allow the $\uD 7$-branes to have multiple windings on the
4-cycles (as we do in the next section), we will not have the freedom
to arrange for the near-alignment of decay constants for the even
axions $\vt^{\alpha}$.  We will therefore focus first on the odd
axions $c^{a}$.\footnote{See, for
  example,~\cite{Grimm:2007hs,McAllister:2008hb,Berg:2009tg,Ben-Dayan:2014zsa,Ben-Dayan:2014lca}
  for other models where these particular axions drive inflation.}

In order to obtain alignment in the odd sector, we need  
$h_{-}^{1,1}\ge 2$.  For simplicity, we will assume that the overall
volume of the Calabi-Yau takes the `strong Swiss cheese' form
\begin{equation}
  \cV=\bigl(\tau^{1}\bigr)^{3/2}-\ga\bigl(\tau^{2}\bigr)^{3/2}\, .
\end{equation} We will further assume that $h_{-}^{1,1}=2$
and that the odd cycles intersect only with $\cD^{2}$, the volume of
which is controlled by $\tau^{2}$.  (In fact our construction allows
for intersections with other cycles, as long as these cycles are not
wrapped by branes that would contribute to the superpotential for
$c^{a}$.)

The other crucial ingredient in the alignment scenario is the
existence of two non-perturbative potentials coupling to different
combinations of the axions $c^{a}$.  Using the results of the previous
section, this can be arranged by taking two stacks of magnetized $\uD
7$-branes wrapping distinct representatives of $\cD^{2}$.
Configurations of this form can arise if the cycle $\cD^{2}$ is not
rigid, so that the D7-branes have deformation moduli in the absence of
flux, but are stabilized by flux on distinct representatives.  An
alternative possibility is that at the geometric level, before the
inclusion of flux, $\cD^{2}$ has more than one isolated locally
volume-minimizing representative.

\begin{figure}
\begin{center}
  \includegraphics[width=\columnwidth]{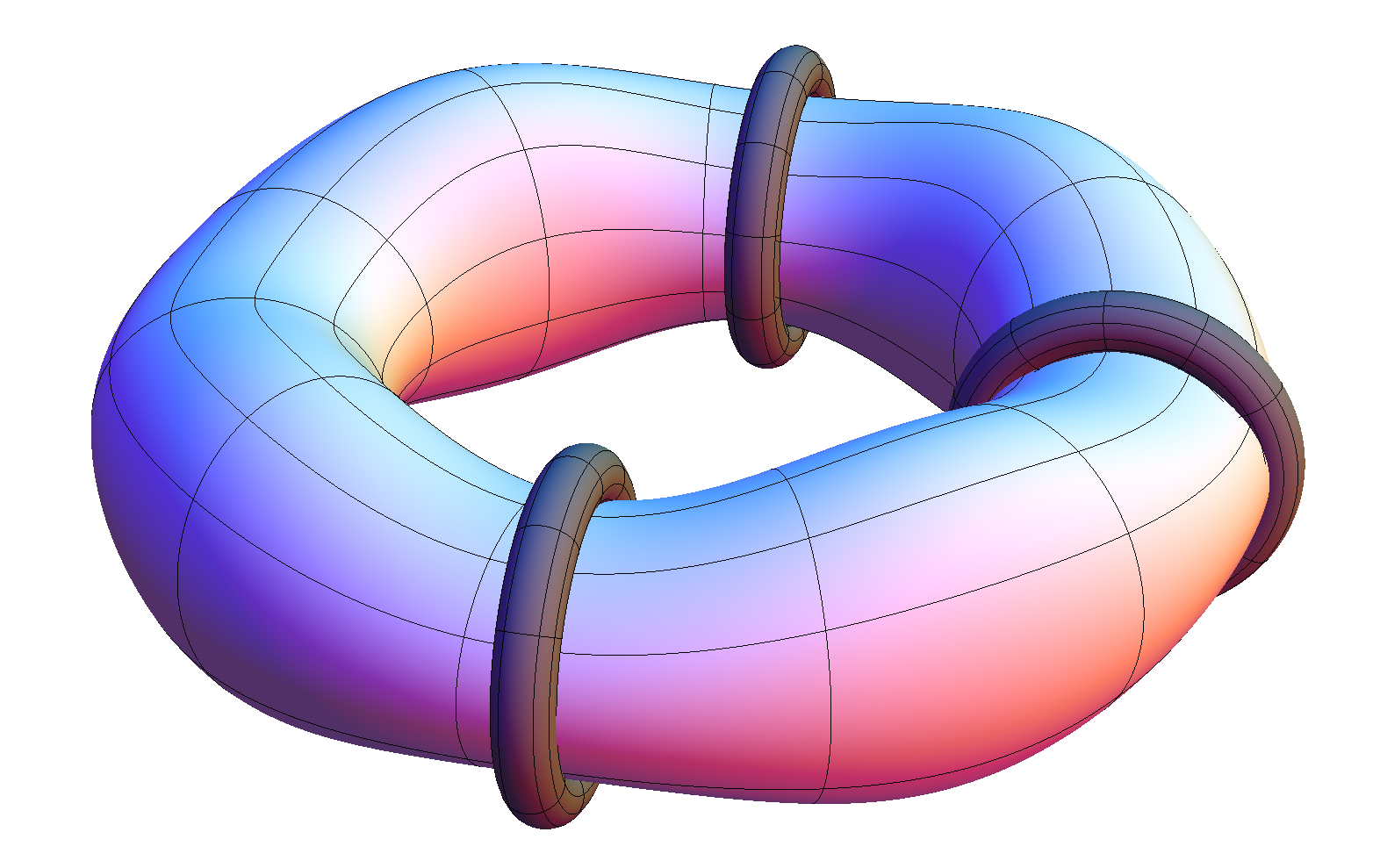}
  \caption{\label{fig:mag}Three branes wrapping locally volume
    minimizing representatives of the same homology class.}
\end{center}
\end{figure}

By an appropriate choice of worldvolume flux on the two stacks, we
will be able to arrange for approximate alignment of the decay
constants, as the gauge kinetic function for the magnetized D7-brane
will depend linearly on the odd moduli~\eqref{eq:gkf}.  However,
$\hat{\vt}^{\alpha}=\mathrm{Im}\,T^{\alpha}$ remains an unstabilized
axion that appears in the same nonperturbative effect.  Since the
appearance of $\hat{\vt}^{\alpha}$ in the gauge kinetic function is
independent of the magnetization, it is difficult to utilize the
$C_{4}$ axions in the scheme of decay constant alignment via
magnetization. We therefore consider a third stack, wrapping a third
distinct representative (see figure~\ref{fig:mag}).  This unmagnetized
stack will allow us to stabilize the $\tau^{\alpha}$,
$\hat{\vt}^{\alpha}$, and $b^{a}$.  In addition, $b^{a}$ and
$\tau^{\alpha}$ will receive contributions to their potentials from
D-terms~\eqref{eq:D-term}.  These ingredients can be used together to
stabilize the saxions $\tau^{\alpha}$ and $b^{a}$, as well as the
additional axion $\hat{\vt}^{\alpha}$, at a higher scale than $c^{a}$.

With these ingredients in hand, the non-perturbative superpotential
from gaugino condensation on the three different stacks takes the form
\begin{equation}
  W_{\np}=\sum_{\xi=A,B,C}A_{\xi}e^{-\frac{2\pi}{N_{\xi}}\mathfrak{f}_{\xi}},
\end{equation}
in which
\begin{equation}
  \mathfrak{f}_{\xi}=T^{2}+i\ka^{2}_{\phantom{2}bc}
  \magf^{b}_{\xi}\bigl(G^{c}+\frac{\tau}{2}\magf^{c}_{\xi}\bigr),
\end{equation}
where $\magf^{a}_{\xi}$ are the worldvolume flux
quanta~\eqref{eq:wv_flux}.  The unmagnetized stack has $\magf^{a}_{C}=0$
and so $\mathfrak{f}_{C}=T^{2}$.  These non-perturbative superpotential terms
supplement the classical flux contribution~\cite{Gukov:1999ya}
\begin{equation}
  W_{0}=\int G_{3}\wedge\Omega.
\end{equation}
The F-term contribution to the scalar potential is
\begin{equation}
  \label{eq:Fterm_pot}
  V_{F}=e^{K}
  \bigl(K^{i\bar{\jmath}}D_{i}W\bar{D}_{\bar{\jmath}}\bar{W}
  -3\left\lvert W\right\rvert^{2}\bigr),
\end{equation}
in which $D_{i}W=W_{,i}+K_{,i}W$ is the K\"ahler-covariant derivative
of the superpotential, $K^{i\bar{\jmath}}$ is the inverse of the
K\"ahler metric $K_{i\bar{\jmath}}=K_{,i\bar{\jmath}}$, and the sum
in~\eqref{eq:Fterm_pot} runs over all chiral fields.  The complex
structure moduli $z$ are assumed to be supersymmetrically stabilized
at a high scale and so we take $D_{z}W=0$.  We assume also that
$\tau^{\alpha}$, $\vt^{\alpha}$, and $b^{a}$ are all stabilized by
D-terms and by the unmagnetized condensate.  We further assume that
$b^{a}=0$ at the minimum, so that the K\"ahler metric for the even and
odd moduli becomes block diagonal, $K_{\alpha\bar{a}}=0$.  Using the
no-scale result (which applies even when $b^{a}\neq 0$)
\begin{equation}
  K^{I\bar{J}}K_{I}K_{\bar{J}}=3,
\end{equation}
where the sum is over all K\"ahler moduli, and assuming that $W_{0}$
is large compared to $W_{\np}$, the scalar potential takes the form
\begin{equation}
  V=-\frac{2}{\cV^{2}}
  \bigl(\tau^{\alpha}\overline{W}_{0}\partial_{\alpha}
  W_{\mathrm{np}}+\mathrm{c.c}\bigr)
  +V_{\mathrm{uplift}},
\end{equation}
in which we have included an uplift potential coming from ingredients
such as anti-D3-branes or frustrated D-terms. By an appropriate tuning of
$V_{\mathrm{uplift}}$, the potential for the $c^{a}$ axions takes the
form necessary for decay constant alignment~\eqref{eq:KNP_pot}
\begin{equation}
  V=\sum_{\xi}\Lambda_{\xi}^{4}
  \biggl[1-\cos\left(\frac{2\pi}{N_{\xi}}\ka^{2}_{\phantom{2}bc}
    \magf^{b}_{\xi}c^{c}\right)\biggr],
\end{equation}
in which
\begin{equation}
  \Lambda_{\xi}^{4}=\frac{8\pi\left\lvert W_{0}\right\rvert A_{\xi}}
         {\cV^{2}N_{\xi}}
  \tau^{2}\,
  e^{-\frac{2\pi}{N_{\xi}}\left(\tau^{2}
    -\frac{1}{2g_{\us}}\ka^{2}_{\phantom{2}bc}
    \magf^{b}_{\xi}\magf^{c}_{\xi}\right)},
\end{equation}
where we have assumed that $W_{0}<0$ and $A_{\xi}>0$ and have set the
stabilized axiodilaton to $\tau=\frac{i}{g_{\us}}$.

With $b^{a}$ stabilized at zero, the metric for the odd 
moduli is
\begin{equation}
  K_{a\bar{b}}=-\frac{g_{\us}}{\cV}\ka^{\alpha}_{\phantom{\alpha}ab}
  t_{\alpha}.
\end{equation}
If we consider an example with $\ka^{2}_{\phantom{2}bc}=-\delta_{bc}$,
then the kinetic term for the odd axions is
\begin{equation}
  \cL^{\mathrm{kin}}=-\frac{g_{\us}t_{2}}{\cV}
  \partial_{\mu}c^{a}\partial^{\mu}c^{a}.
\end{equation}
The determinant of the Hessian at the minimum is
\begin{equation}
  \label{eq:mag_Hess}
  \left.\det \partial_{a}\partial_{b}V\right\rvert_{c=0}
  =\frac{\left(f_{1}^{A}f_{2}^{B}-f_{1}^{B}f_{2}^{A}\right)^{2}}
  {N_{A}^{2}N_{B}^{2}}\frac{2\pi^{4}\cV\Lambda_{A}^{4}\Lambda_{B}^{4}}{t_{2}g_{\us}}.
\end{equation}
The form of~\eqref{eq:mag_Hess} exhibits the effect of decay constant
alignment~\eqref{eq:KNP_Hess_det}.  If we relax the assumption
$\ka^{2}_{\phantom{2}bc}=-\delta_{bc}$, then the resulting kinetic
mixing will generally make the task of arranging a suitably flat
potential more delicate.

\newpage

%%%%%
%%%%%
\section{\label{sec:winding}Large Decay Constants from Multiple Windings}
%%%%%
%%%%%

In the example of the previous section, we made use of the $C_{2}$
axions $c^{a}$ since their decay constants could be aligned by
adjusting the magnetic flux on the $\uD 7$-branes. While the decay
constants for the $C_{4}$ axions $\vt^{\alpha}$ cannot be altered by
magnetization, they can be aligned by winding numbers if we allow
multiply-wrapped $\uD 7$-branes.

To realize alignment for $\vt^{\alpha}$, we need 
$h_{+}^{1,1}\ge 2$. For simplicity, we consider examples with
$h_{-}^{1,1}=0$ (in which case $\hat{\vt}^{\alpha}=\vt^{\alpha}$). We
furthermore need two separate condensates to generate two distinct
non-perturbative terms in the superpotential.  As in the previous
section, this can be arranged by taking two stacks of $\uD 7$-branes,
each of which wraps a different representative of the two homology
classes. The non-perturbative superpotential realized on these
condensates is
\begin{equation}
  W_{\np}=A_{A}e^{-\frac{2\pi}{N_{A}}\left(T^{1}+T^{2}\right)}
    +A_{B}e^{-\frac{2\pi}{N_{B}}\left(T^{1}+T^{2}\right)}.
\end{equation}

\begin{figure}
\begin{center}
  \includegraphics[width=\columnwidth]{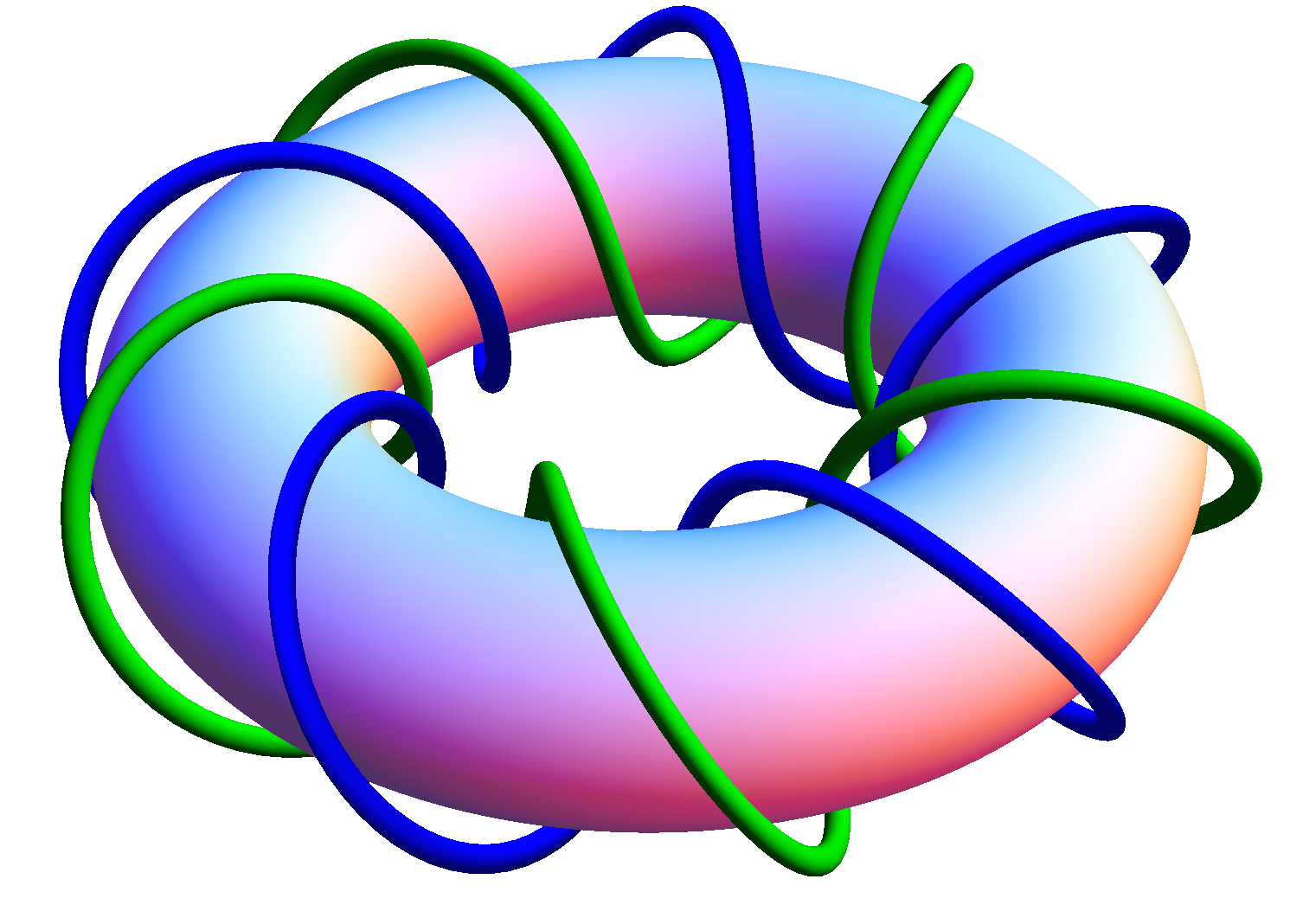}
  \caption{\label{fig:winding}Two branes multiply wrapping the same
    cycle the same number of times.}
\end{center}
\end{figure}

The fact that both $T^{1}$ and $T^{2}$ appear in the same linear
combination eliminates our ability to arrange for near alignment of
the decay constants $\vt^{\alpha}$.  However, we gain additional
freedom in aligning decay constants by allowing the D7-branes to wrap
multiple times.\footnote{A critical assumption is that the separate
  windings are spaced apart, as illustrated in
  figure~\ref{fig:winding}, so that there are no new light non-abelian
  gauge bosons from the multi-wrappings.  Arranging for spiraling
  wrappings of this form may be possible given suitable fluxes, as we
  discuss below.} Including such wrapping numbers modifies the
non-perturbative superpotential~\eqref{eq:gkf} to
\begin{equation}
  W_{\np}=A_{A}e^{-\frac{2\pi}{N_{A}}
    \left(\windnum_{1}^{A}T^{1}+\windnum_{2}^{A}T^{2}\right)}
    +A_{B}e^{-\frac{2\pi}{N_{B}}
    \left(\windnum_{1}^{B}T^{1}+\windnum_{2}^{B}T^{2}\right)}.
\end{equation}
These multiple windings effectively increase the volume of the
$\uD 7$-branes while keeping the rank of the low-energy gauge group
unchanged. Again, by a suitable tuning of $V_{\mathrm{uplift}}$, we
have a potential of the form~\eqref{eq:KNP_pot}
\begin{align}
  V=\sum_{\xi=a,b}\Lambda_{\xi}^{4}
  \biggl\{1-\cos\biggl[\frac{2\pi}{N_{\xi}}
    \bigl(\windnum_{1}^{\xi}\vt^{1}+\windnum_{2}^{\xi}\vt^{2}\bigr)\biggr]\biggr\},
\end{align}
in which
\begin{equation}
  \Lambda_{\xi}^{4}
  =\frac{8\pi \left\lvert W_{0}\right\rvert A_{\xi}}{N_{\xi}\cV^{2}}
  \biggl(\windnum_{1}^{\xi}\tau^{1}+\windnum_{2}^{\xi}\tau^{2}\biggr)
  e^{-\frac{2\pi}{N_{\xi}}\left(\windnum_{1}^{\xi}\tau^{1}
    +\windnum_{2}^{\xi}\tau^{2}\right)},
\end{equation}
where we have taken $-W_{0}$ and $A_{\xi}$ to be real and positive.

Although the number of $\uD 7$-branes $N_{\xi}$ and the winding
numbers $\windnum_{\alpha}^{\xi}$ are integers, they provide enough freedom
to realize the near alignment of the axion decay constants.  However,
the story is complicated by the fact that the scalar fields
$T^{\alpha}$ do not have canonical kinetic terms and will be subject
to kinetic mixing.  As a simple example, we assume that the volume
takes the form
\begin{equation}
  \cV=\left(\tau^{1}\right)^{3/2}-\ga\left(\tau^{2}\right)^{3/2},
\end{equation}
for some $\ga>0$.  The kinetic term for the axions is then
\begin{equation}
  \cL^{\mathrm{kin}}=-g_{\alpha\beta}\partial_{\mu}\vt^{\alpha}\partial^{\mu}\vt^{\beta},
\end{equation}
with
\begin{equation}
  g_{ij}=\frac{1}{\cV^{2}}
  \begin{pmatrix}
    \frac{6\left(\tau^{1}\right)^{3/2}+3 \ga\,\left(\tau^{2}\right)^{3/2}}
    {8\sqrt{\tau^{1}}} &
    -\frac{9\ga\sqrt{\tau^{1}\tau^{2}}}{8} \\
    -\frac{9\ga\sqrt{\tau^{1}\tau^{2}}}{8} &
    \frac{3\ga \left(\tau^{1}\right)^{3/2}+6 \ga^{2}\,\left(\tau^{2}\right)^{3/2}}
    {8\sqrt{\tau^{2}}}
  \end{pmatrix}.
\end{equation}
The determinant of the Hessian at the minimum of the potential is
(after canonically normalizing the fields)
\begin{multline}
  \left.\det \partial_{\alpha}\partial_{\beta}V\right\rvert_{\vt=0}\\
  =\frac{\left(\windnum_{1}^{A}\windnum_{2}^{B}-\windnum_{1}^{B}\windnum_{2}^{A}\right)^{2}}{N_{A}^{2}N_{B}^{2}}
  \frac{128\pi^{4}\cV^{2}\sqrt{\tau^{1}\tau^{2}}\Lambda_{B}^{4}\Lambda_{B}^{4}}
  {9\ga}.
\end{multline}
Again, this form exhibits the flat direction that appears upon
alignment of the decay constants~\eqref{eq:KNP_Hess_det}.

%%%%
%%%%
\section{\label{sec:examples}Examples}
%%%%
%%%%

We can illustrate the parametric success of alignment in these
constructions by considering some particular examples. Although we
provide only a few toy cases here, the mechanism to arrange for decay
constant alignment with our ingredients is more general.
Further control could be obtained by using additional
stacks and axions to align multiple decay constants as in the
field-theoretic treatment of~\cite{Choi:2014rja}.

\subsection{Alignment from multiply wrapped branes}

\begin{figure}
\begin{center}
  \includegraphics[width=\columnwidth]{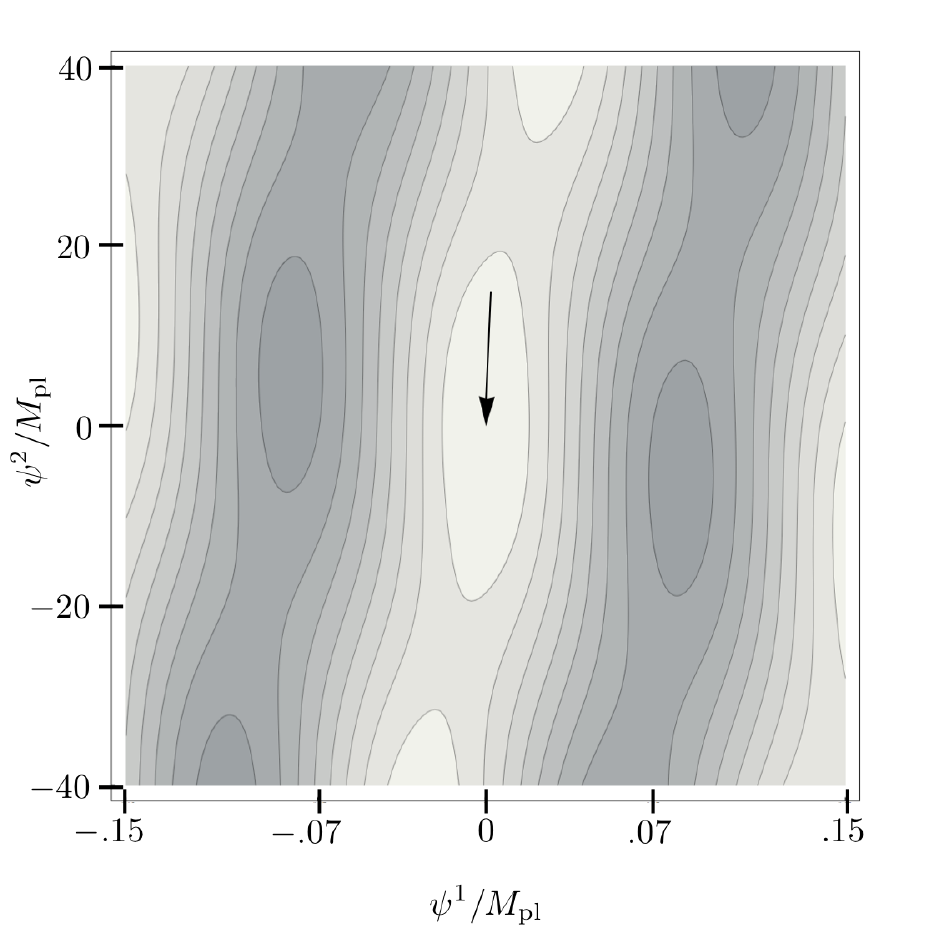}
  \caption{\label{fig:potential}The 2-axion
    potential~\eqref{eq:example_pot}.  Darker regions correspond to
    higher points in the potential. Inflation can be accommodated
    along the super-Planckian trajectory indicated by the arrow.  The
    very slight tilt of the trajectory is a consequence of a very
    slight shift in the position of the minimum for $\psi^{1}$ as
    $\psi^{2}$ evolves.}
\end{center}
\end{figure}

We first present an example of alignment resulting from multiply
wrapped branes discussed in \S\ref{sec:winding}. We consider a
strong Swiss cheese Calabi-Yau such that the volume can be written as
\begin{equation}
  \label{eq:wound_vol}
  \cV=\bigl(\tau^{1}\bigr)^{3/2}-\bigl(\tau^{2}\bigr)^{3/2}.
\end{equation}
We take this toy example just to illustrate the success of alignment,
and the scheme will work more generally in real Calabi-Yaus provided that
multiple windings can be accommodated. We wrap two stacks of
$\uD 7$-branes with the following data for stack A:
\begin{equation}
  N_{A}=35,\quad
  \windnum_{1}^{A}=1,\quad \windnum_{2}^{A}=19,
\end{equation}
while for stack B:
\begin{equation}
  N_{B}=30,\quad
  \windnum_{1}^{B}=1,\quad \windnum_{2}^{B}=20.
\end{equation}
Taking
\begin{equation}
  -W_{0}=A_{A}=A_{B}=.1,
\end{equation}
and considering the point $\tau^{1}=15$, $\tau^{2}=2$, the masses at the
minimum are
\begin{equation}
  \label{eq:example_masses}
  m_{1}=4\times 10^{-3}\Mpl,\quad
  m_{2}=4\times 10^{-6}\Mpl.
\end{equation}
In terms of these mass eigenstates, the potential takes the form
\begin{align}
  \frac{V}{\Mpl^{4}}\approx&\left(1.1\times 10^{-8}\right)\notag\\
  &-\left(9.2\times 10^{-9}\right)\cos
  \left[-\frac{37\,\psi^{1}}{\Mpl}+\frac{0.02\,\psi^{2}}{\Mpl}\right]
  \label{eq:example_pot}\\
  &-\left(1.5\times 10^{-9}\right)\cos
  \left[\frac{46\,\psi^{1}}{\Mpl}+\frac{.099\,\psi^{2}}{\Mpl}\right].\notag
\end{align}
Thus, $\psi^{1}$ can be consistently integrated out, and what remains is a
potential for $\psi^{2}$ that can accommodate natural inflation (see
figure~\ref{fig:potential}).  Note that the mass for $\psi^{2}$
in~\eqref{eq:example_masses} is the appropriate scale for realizing
$m^2\phi^2$ chaotic inflation.

The large ranks of the $\uD 7$-branes and the relatively small
$\tau^{2}$ also lead to parametric increases in the decay constant, so
we should ensure that the enhancements in the decay constants are in
fact a result of the alignment and not of the large rank or small
4-cycle volume.  To see this, we could repeat the analysis of the
above example but remove the possibility for alignment by taking
$\windnum_{1}^{A}=\windnum_{2}^{B}=1$ and
$\windnum_{1}^{B}=\windnum_{2}^{A}=0$ (note that increasing the
winding numbers decreases the decay constants).  In this reference
case, the potential in the mass eigenbasis is
\begin{align}
  \frac{V}{\Mpl^{4}}\approx &\left(6\times
  10^{-6}\right)\notag\\ &-\left(2\times
  10^{-6}\right)\cos\biggl[-\frac{1.5\psi^{1}}{\Mpl}
    +\frac{1.7\psi^{2}}{\Mpl}\biggr]\\ &-\left(4\times
  10^{-6}\right)\cos
  \biggl[\frac{2.1\psi^{1}}{\Mpl}+\frac{.6\psi^{2}}{\Mpl}\biggr].
  \notag
\end{align}
Although the decay constants are indeed relatively large, they are
only $\cO\left(\Mpl\right)$, and natural inflation could not be
sustained.  Further increasing the number of $\uD 7$-branes so that
the total tension of the $\uD 7$-branes is comparable to the total
tension of the multiply wound branes would increase the decay
constants, but would also suppress the dynamical scales, making it
difficult to match the COBE normalization of the scalar power
spectrum.

The example above exists at relatively small volume, and $\alpha'$
corrections are of concern.  We can move to a point of larger volume,
but this acts to suppress the size of the non-perturbative effects,
which can be compensated by an increase in the ranks of the condensing
gauge groups.  For example, if we consider the point $\tau^{1}=25$,
$\tau^{2}=10$, $N_{A}=100$, $N_{B}=90$,
$\windnum_{1}^{A}=\windnum_{1}^{B}=1$, $\windnum_{2}^{B}=20$, and
$\windnum_{2}^{A}=19$ with $-W_{0}=10$, $A_{A}=A_{B}=1$ we find the
potential
\begin{align}
  \frac{V}{\Mpl^{4}}\approx &\left(9.5\times
  10^{-8}\right)\notag\\ &-\left(8.4\times
  10^{-8}\right)\cos\biggl[-\frac{30\psi^{1}}{\Mpl}
    +\frac{.0065\psi^{2}}{\Mpl}\biggr]\\ &-\left(1.1\times
  10^{-8}\right)\cos
  \biggl[-\frac{35\psi^{1}}{\Mpl}+\frac{.04\psi^{2}}{\Mpl}\biggr].
  \notag
\end{align}
This again supports natural inflation.  Once again, the large decay
constants are not a result of the large ranks, but instead a
manifestation of the power of decay constant alignment.  Setting
$\windnum_{1}^{A}=\windnum_{2}^{B}=1$ and
$\windnum_{1}^{B}=\windnum_{2}^{A}=0$, we again find
$\cO\left(\Mpl\right)$ decay constants (which are again larger than
one expects at generic points, but are not large enough to sustain
inflation).

In either of these cases, obtaining realistic inflation using just the
non-perturbative superpotentials requires a non-perturbative effect
that is not too small.  Since the additional windings increase the
action of the $\uD 7$-branes, arranging for alignment quickly pulls
down the dynamical scale unless the rank of the gauge groups is
increased or the volume of the cycles is shrunk.  This is an important
constraint on realizing inflation via decay constant alignment that
cannot be seen in field theoretic treatments where the decay constants
and dynamical scales are independently adjustable.  If we are not
concerned with realizing the COBE normalization, then we can obtain
large decay constants with more moderate volumes and ranks.  For
example, if we take $\tau^{1}=25$, $\tau^{2}=10$, $N_{A}=N_{B}=15$,
$\windnum_{1}^{A}=\windnum_{1}^{B}=1$, $\windnum_{2}^{B}=30$, and
$\windnum_{2}^{A}=29$ with $-W_{0}=A_{A}=A_{B}=1.0$ we find a decay
constant of $\sim 5\,\Mpl$ (compared to $.2\Mpl$ without the mixing),
but the mass of the would-be inflaton is
$\cO\left(10^{-31}\Mpl\right)$, which is far too small to produce the
observed CMB anisotropies.  This could still be useful for
constructing models of axion monodromy
inflation~\cite{Silverstein:2008sg,McAllister:2008hb} in which the
shift symmetry of the axion plays an important role, but inflation is
driven by explicit breaking terms.

\subsection{Alignment from magnetized branes}

As a final example we  present a realization of the magnetized
brane scenario  supporting natural inflation.
We write the volume of the Calabi-Yau in a form similar to that of the
previous examples~\eqref{eq:wound_vol}
\begin{equation}
\mathcal{V} = \frac{1}{6}\left( t_1^3 - t_2^3 \right),
\end{equation}
where $t_{1}$ and $t_{2}$ control the volumes of 2-cycles and are
constrained to be positive.  The $\uD 7$-branes are taken to wrap a
divisor whose volume is $\tau^{2}=\frac{1}{2}t_{2}^{2}$. For
simplicity, we assume that the only non-vanishing even-odd-odd
intersection numbers are of the form
\[ \kappa^2_{\ ab} = \begin{pmatrix}
  -2 & 1 \\
  1 & -2 \end{pmatrix}.
\]
As in the previous case, the intersection numbers that we have chosen
in this example are just for illustrative purposes and the mechanism
will work more generally. We wrap two stacks of magnetized D7-branes,
with the following data for stack A:
\begin{equation}
  N_{A}=35,\quad
  \magf_{1}^{A}=1,\quad \magf_{2}^{A}=6,
\end{equation}
while for stack B:
\begin{equation}
  N_{B}=40,\quad
  \magf_{1}^{B}=1,\quad \magf_{2}^{B}=7.
\end{equation}
Taking
\begin{equation}
  -W_{0}=A_{A}=A_{B}=1, \quad
  g_s = 0.5,
\end{equation}
and considering the point $t_{1}=7$, $t_{2}=3$, the masses at the
minimum are
\begin{equation}
  \label{eq:example_masses2}
  m_{1}=4\times 10^{-4}\Mpl,\quad
  m_{2}=2\times 10^{-6}\Mpl.
\end{equation}
In terms of these mass eigenstates, the potential takes the form
\begin{align}
  \frac{V}{\Mpl^{4}}\approx&\left(8.3\times 10^{-9}\right)\notag\\
  &-\left(7.6\times 10^{-9}\right)\cos
  \left[\frac{4\,\psi^{1}}{\Mpl}+\frac{0.006\,\psi^{2}}{\Mpl}\right]
  \label{eq:example_pot2}\\
  &-\left(6\times 10^{-10}\right)\cos
  \left[-\frac{4\,\psi^{1}}{\Mpl}+\frac{.066\,\psi^{2}}{\Mpl}\right],\notag
\end{align}
which again supports natural inflation.

%%%%%
%%%%%
\section{Conclusions}
%%%%%
%%%%%

In this note, we have presented an embedding into type IIB  string theory of 
the field-theoretic axion decay constant alignment mechanism proposed by
Kim, Nilles,
and Peloso~\cite{Kim:2004rp}. Our primary tool is gaugino
condensation on multiple stacks of $\uD 7$-branes wrapping homologous
cycles in a Calabi-Yau orientifold.  When the branes are magnetized,
gaugino condensation leads to non-perturbative superpotentials that
give the  leading breaking of the shift symmetry of the $C_{2}$
axions.  Approximate alignment of the decay constants can then be achieved
by an appropriate choice of the magnetic fluxes.  Alternatively, if
the $\uD 7$-branes are multiply wound, their couplings to
$C_{4}$ axions can be aligned by adjusting the winding numbers.

Although our constructions require only ingredients that are
commonplace in stabilized flux vacua of type IIB string theory
compactified on O3/O7 orientifolds of Calabi-Yau manifolds, further
care must be taken to ensure compatibility of moduli stabilization
with axion alignment.  Consistently stabilizing all moduli, leading to
a fully-realized model of large-field inflation in string theory,
remains a significant challenge in our construction, just as in all
alternative scenarios for inflation in string theory.  In the
closed-string sector, this may be particularly delicate in the
magnetized case of \S\ref{sec:mag} where the $B_{2}$ saxions $b^{a}$
must have small vevs to ensure that mixing between the even and odd
moduli can be neglected and that the dynamical scales (which are
exponentially sensitive to $b^{a}b^{b}/g_s$) are not too suppressed.
Although the case of winding branes of \S\ref{sec:winding} does not
suffer from such a dramatic saxion problem (though of course the
4-cycle volumes must be stabilized), it may be difficult to arrange
for the required winding numbers without the $\uD 7$-branes
intersecting themselves or each other.  Such intersections will
introduce additional vector-like matter fields that must be made
massive in order for the non-perturbative effects that our
constructions invoke to be present.  Even if such intersections can be
avoided, finding volume-minimizing cycles allowing for multiple
windings, or fluxes that stabilize such windings, may be
difficult. However, once arranged, the masses of the $\uD 7$-brane
moduli will be comparable to the masses of the complex structure
moduli and so these fields will be inert during inflation.  An
additional constraint is that the $\uD 7$-brane charge (and induced
lower brane charge in the magnetized case) must be canceled.  Although
such cancellation can be achieved by orientifold planes, it requires
more detailed constructions than those that we provide here.

Finally, in schemes of decay constant alignment, there is some tension
between the arrangement of a large effective decay constant and the
scale needed to match the normalization of the scalar power spectrum.
This can be seen in \eqref{eq:KNP_pot}, where although the decay
constant is indeed enhanced by the misalignment $\delta$, it is at a
cost of a decreased inflaton mass.  In string-theoretic
implementations, this problem is exacerbated as the dynamical scale
will often decrease exponentially as the misalignment is obtained.
This tension is not a fatal flaw of our proposal, and indeed it is
easy to find examples with properly normalized scalar power spectra.

Some of these difficulties can be ameliorated by combining our scheme
with other recent proposals for producing large effective decay
constants.  In particular, in~\cite{Choi:2014rja} it was shown that
the success of decay constant alignment in field-theoretic models can
be extended by increasing the number of axions that mix in the scalar
potential.  Chaining together alignment effects would be useful in our
construction, because although alignment does provide for a parametric
enhancement, the points in moduli space that exhibit natural inflation
are often near the edge of control.  Using a combination of alignments
would allow us to obtain parametrically large axion decay constants
within a region of robust control.  In addition, it was recently
demonstrated~\cite{Bachlechner:2014hsa} that the kinematic extension
of field range resulting from the combination of many axions
{\it{without}} aligned decay constants can be more efficient than
suggested previously in the N-flation
literature~\cite{Dimopoulos:2005ac}. The kinetic alignment effect
of~\cite{Bachlechner:2014hsa} in concert with decay constant alignment
may be particularly powerful.  Moreover, one could use axion alignment
to build a broader range of scenarios for axion
monodromy~\cite{Silverstein:2008sg,McAllister:2008hb}: see in
particular the two-axion monodromy constructions
of~\cite{Berg:2009tg}.

Constructing a fully-stabilized compactification that implements our proposal  
is an important problem for the future.

%%%%
%%%%
%\acknowledgements
\section*{Acknowledgements}
%%%%
%%%%

We thank Thomas Bachlechner, John Stout, and Timm Wrase for related
discussions.  We are also grateful to Ido Ben-Dayan, Francisco Pedro,
and Alexander Westphal for sharing a draft of their
forthcoming related work.  We are especially indebted to Julia
Goodrich for technical assistance.  This work was supported by NSF
grant PHY-0757868.

\bibliography{lmm}

\end{document}